\documentclass[]{sig-alternate-2013}
\allowdisplaybreaks
\usepackage{cite,url}
\usepackage{booktabs}
\usepackage{amsmath,amssymb,bm}
\usepackage[dvips]{color}
\usepackage{graphicx}

\usepackage[lined,boxed,commentsnumbered, ruled]{algorithm2e}
\usepackage{booktabs}
\usepackage{subfigure}
\usepackage{amsfonts}
\newdef{theorem}{Theorem}
\newdef{lemma}{Lemma}
\newdef{definition}{Definition}
\newdef{assumption}{Assumption}
\newdef{corollary}{Corollary}
\newdef{remark}{Remark}

\newcommand {\bx}{\mathbf{x}}
\newcommand {\by}{\mathbf{y}}
\newcommand {\bzero}{\mathbf{0}}

\makeatletter
\let\@copyrightspace\relax
\makeatother

\begin{document}


\title{Understanding the Benefits of Open Access in Femtocell Networks: Stochastic Geometric Analysis in the Uplink }

\author{
\alignauthor
Wei Bao and Ben Liang\\
Department of Electrical and Computer Engineering, University of Toronto\\
       \email{\{wbao,liang\}@comm.utoronto.ca}
}

\maketitle

\begin{abstract}
We introduce a comprehensive analytical framework to compare between open access and closed access in two-tier femtocell networks, with regard to uplink interference and outage.  Interference at both the macrocell and femtocell levels is considered.  A stochastic geometric approach is employed as the basis for our analysis.  We further derive sufficient conditions for open access and closed access to outperform each other in terms of the outage probability, leading to closed-form expressions to upper and lower bound the difference in the targeted received power between the two access modes.  Simulations are conducted to validate the accuracy of the analytical model and the correctness of the bounds.  
\end{abstract}



\section{Introduction}\label{section_intro}

In deploying wireless celluar networks, some of the most important objectives are to provide higher capacity, better service quality, lower power usage, and ubiquitous coverage.  To achieve these goals, one efficient approach is to install a second tier of smaller cells, which are referred to as femtocells, overlapping the original macrocell  network \cite{intro}. Each femtocell is equipped with a short-range and low-cost base station (BS).

In the presence of femtocells, whenever a User Equipment (UE) is near a femtocell BS, two different access mechanisms may be applied: closed access and open access. Under closed access, a femtocell BS only provides service to its local users, without further admitting nearby macrocell users. In contrast, under open access, all nearby macrocell users are allowed to access the femtocell BS.  The open access mode increases the interference level from within a femtocell, but it also allows macrocell UEs that might otherwise transmit at a high power toward their faraway macrocell BS to potentially switch to lower-power transmission toward the femtocell BS, therefore reducing the overall interference in the system.  However, the relative merits between open access and closed access remain unresolved within the research community, as they may concern diverse factors in communication efficiency, control overhead, system security, and regulatory policies.

In this work, we contribute to the current debate by providing new technical insights on how the two access modes may affect both macrocell users and local femtocell users, in terms of the \textit{uplink} interference and outage probabilities.  We seek to quantify the conditions to guarantee that one access mode improves the performance of macrocell or femtocell users.  It is a challenging task, as we need to account for the diverse spatial patterns of different network components.  Macrocell BSs are usually deployed regularly by the network operator, while femtocell BSs are spread irregularly, sometimes in an anywhere plug-and-play manner, leading to a high level of spatial randomness.  Furthermore, macrocell users are randomly distributed throughout the system, while femtocell users show strong spatial locality and correlation, since they aggregate around femtocell BSs.  Whenever open access is applied, we also need to consider the effects of handoffs made by open access users, which brings even more complication to the analytical model.

We develop stochastic geometric analysis schemes to derive numerical expressions for the uplink interference and outage probabilities of open access and closed access by modeling macrocell BSs as a regular grid, macrocell UEs as a Poission point process (PPP), and femtocell UEs as a two-level clustered Poisson point process, which captures the spatial patterns of different network components.  However, uplink interference analysis is notoriously complex even for traditional single-tier cellular networks.  For the two-tier network under consideration, our analysis yields non-closed forms requiring numerical integrations.  This motivates us to further develop closed-form sufficient conditions for open access and closed access to outperform each other, at both the macrocell and femtocell levels.

Based on the above analysis, we are able to extract a key factor that influences the performance difference between open access and closed access: the power enhancement factor $\rho$, which is the ratio of the targeted received power of an open access user to its original targeted received power in the macrocell.
We investigate the threshold value $\rho^{*}$ (resp.~$\rho^{**}$) such that
macrocell (resp.~femtocell) users may benefit through open access if  $\rho<\rho^{*}$ (resp.~$\rho<\rho^{**}$)  as we apply open access to replace closed access.  Tight upper and lower bounds of $\rho^{*}$ are derived in closed forms, and the bounds of $\rho^{**}$ can be found by numerically searching through a closed-form equation, providing system design guidelines with low computational complexity.
To the best of our knowledge, this is the first paper to theoretically analyze the uplink performance difference between open access and closed access of femtocell networks that considers the impact of random spatial patterns of BSs and UEs.

The rest of the paper is organized as follows: In Section \ref{section_related}, we discuss the relation between our work and prior works.
In Section \ref{section_model}, we present the system model. 
In Section \ref{section_tier1} and \ref{section_tier2}, we analyze the performance at the macrocell and femtocll levels, respectively. In Section \ref{section_numerical}, we validate our analysis with simulation results.
 Finally, concluding remarks are given in Section \ref{section_conclusion}.

\section{related works} \label{section_related}

The downlink interference and outage performance in cellular networks have been  extensively studied using the stochastic geometric approach. \cite{SG_MultiTier0, SG_MultiTier} analyzed the downlink performance of heterogeneous networks with multiple tiers by
assuming the signal-to-interference plus noise ratio (SINR) threshold is greater than $1$. \cite{Hetero_downlink2}  studied the maximum tier-1 user and tier-2 cell densities under downlink outage constraints. \cite{SG_down_load} studied the downlink interference considering load balance. \cite{Hetero_downlink3} studied the downlink user achievable rate in a heterogeneous network considering both SINR and spatial user distributions. \cite{arXiv_downlink} studied  open access versus closed access in femtocell networks in terms of downlink performance.

The analysis of uplink interference in multi-tier networks is more challenging compared with the downlink case.
For uplink analysis, the interference generators are the set of UEs, which are more complicatedly distributed compared with the interference generators (i.e., BSs) in downlink analysis. Under closed access,  without considering random spatial patterns,
 \cite{SG_MultiTier_nonSG1} studied the uplink performance of a single tier-1 cell and a single femtocell, while \cite{SG_MultiTier_nonSG2} extended it to the case of multiple tier-1 cells and multiple femtocells.  \cite{ACM_femtocell1} studied the co-channel uplink interference in LTE-based multi-tier cellular networks, considering a constant number of femtocells in a macrocell. However, none of  \cite{SG_MultiTier_nonSG1, SG_MultiTier_nonSG2, ACM_femtocell1} considered the random spatial patterns of users or femtocells.

By considering random spatial patterns,  \cite{arXiv_uplink} analyzed uplink performance of cellular networks, but it was limited to the one-tier case.
\cite{CDMA_Uplink} evaluated the uplink performance of two-tier networks considering random spatial patterns.
However, several interference components were analyzed based on approximations, such as (1) BSs see a femtocell as a point interference source and (2) Femtocell UEs transmit at the maximum power at the edge of cells.
\cite{SG_ICASSP} studied both uplink and downlink interference of femtocell networks based on a Neyman-Scott Process. However it assumed that each UE transmits at the same power and femtocell users are uniformly distributed in an infinitesimally thin ring around the femtocell BS.  With a more general system model, \cite{OurICCC} derived the uplink interference in a two-tier network with multiple types of users and small cell BSs, but no closed-form result was obtained. Moreover, both \cite{OurICCC, CDMA_Uplink, SG_ICASSP} considered only the closed access case.

The analysis of open access in femtocell networks is even more complicated. This is because the model for open access needs to capture the impact of the users disconnecting from the original macrocell BS and connecting to a femtocll BS. In order to satisfy mathematical tractability, the previous analysis of open access was based on simplified assumptions.  \cite{CDMA_OpenClose} compared the performance of open access and closed access based on a model with one macorcell, one femtocell, and a given number of macrocell users, while \cite{Open2} was based on a model with one macorcell, a constant number of macrocell users, and randomly distributed femtocells. Although \cite{CDMA_OpenClose} and \cite{Open2} provide useful insights into the performance comparison between open access and closed access, due to their limited system models, they have not addressed the challenging issues brought by the diverse spatial patterns of BSs and UEs.


Finally, several other works studied the performance of femtocells based on experiments \cite{ACM_femtocell2, ACM_femtocell3}, which provided important practical knowledge in designing a real system. Compared with these works, our theoretical approach is an essential alternative that allows more rigorous reasoning to understand the performance benefits of open access compared with closed access, by considering more general system models and behaviors instead of specific experimental scenarios.


\section{System Model}\label{section_model}

\begin{figure}[tbp]
\centering  \vspace*{0pt}
\includegraphics[scale=.45]{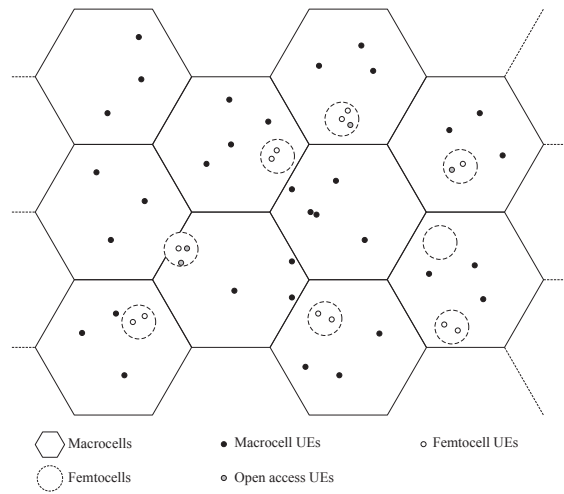}
\caption{Two-tier network with macrocells and femtocells.}
\label{Figmodel}
\end{figure}
\subsection{Two-Tier Network} \label{subsection_topology}
We consider a two-tier network with macrocells and femtocells as shown in Fig.~\ref{Figmodel}. Following the convention
in literature, we assume that the macrocells form an infinite hexagonal grid in the two-dimensional Euclidean space $\mathbb{R}^2$.
Macrocell BSs are located at the centers of the hexagons $\mathbb{B}=\{(\frac{3}{2}aR_c, \frac{\sqrt{3}}{2}aR_c+\sqrt{3}bR_c)|a,b \in\mathbb{Z}\}$, where $R_c$  is the radius of the hexagon. Macrocell UEs are randomly distributed in the system, which are  modeled as a homogeneous Poisson point process (PPP) $\Phi$ with intensity  $\lambda$.

 Because femtocell BSs are operated in a plug-and-play fashion, inducing a high level of spatial randomness,  we assume femtocell BSs  form a homogeneous PPP $\Theta$ with intensity $\mu$. Each femtocell BS is connected to the core network by high-capacity wired links that has no influence on our wireless performance analysis.

Each femtocell BS communicates with local femtocell UEs surrounding it, constituting a femtocell. We assume $R$ as the communication radius of each femtocell BS.
Given the location of a femtocell BS at $\bx_0$, we assume that its femtocell UEs, denoted by  $\Psi(\bx_0)$, are distributed as a non-homogenous PPP in the disk centered at $\bx_0$ with radius $R$. Its intensity at $\bx$ is described by $\nu(\bx-\bx_0)$, a non-negative function of the \textit{vector} $\bx-\bx_0$. Note that  the user intensity   $\nu(\bx-\bx_0)=0$ if $|\bx-\bx_0|>R$. The femtocell UEs in one femtocell are independent with femtocell UEs in other femtocells, as well as the macrocell UEs. We assume the scale of femtocells is much small than the scale of macrocells \cite{intro}, $R\ll R_c$.

  To better understand the spatial distribution of femtocell BSs and femtocell UEs, the femtocell BSs $\Theta$  can be regarded as a parent point process in $\mathbb{R}^2$, while femtocell UEs $\Psi$  is a daughter process associated with a point in the parent point process, forming a two-level random pattern.
  Note that the aggregating of femtocell UEs around a femtocell BS implicitly defines the location correlation among femtocell UEs.




Let $\mathcal{H}(\bx)$ denote the hexagon region centered at $\bx$ with radius $R_c$; let $\mathcal{B}(\bx,R)$ denote the disk region centered at $\bx$ with radius $R$; let $\mathcal{BS}(\bx)$ denote the hexagon center nearest to $\bx$ (i.e., $\mathcal{BS}(\bx)=\bx_0$ $\Leftrightarrow$ $\bx\in\mathcal{H}(\bx_0)$). 

\subsection{Open Access versus Closed Access}

If a macrocell UE is covered by a femtocell BS (i.e., within a distance of $R$ from a  femtocell BS), under closed access, the UE still connects to the macrocell BS. Under open access, the  UE is handed-off to connect to the femtocell BS and \emph{disconnects} from the original macrocell BS; the UE is then referred to as an \emph{open access UE}.

Given a femtocell BS located at $\bx_0$, let $\Omega(\bx_0)$ denote the point process corresponding to the  open access  UEs connecting to it. Note that because the radius of a femtocell is much smaller than that of macrocells, the probability of two femtocells overlapping is small. Thus, $\Omega(\bx_0)$ corresponds  to points of $\Phi$ inside the range of the femtocell BS at $\bx_0$, which is a  PPP  with intensity $\lambda$ inside $\mathcal{B}(\bx_0,R)$.

%
%
%

\subsection{Pathloss  and Power Control}
Let $P_t(\bx)$ denote the transmission power at $\bx$ and $P_r(\by)$ denote the received power at $\by$. We assume that $P_r(\by)=\frac{P_t(\bx)h_{\bx,\by}}{A|\bx-\by|^{\gamma}}$, where $A|\bx-\by|^{\gamma}$ is the propagation loss function with predetermined constants $A$ and $\gamma$ (where $\gamma>2$ in practice),  and $h_{\bx,\by}$ is the fast fading term.  Corresponding to common Rayleigh fading with power normalization, $h_{\bx,\by}$ is independently  exponentially distributed with unit mean.
Let $H(\cdot)$ be the cumulative distribution function of $h_{\bx,\by}$.

We follow the conventional assumption that uplink  power control adjusts for propagation losses  \cite{CDMA_Spatial, CDMA_Uplink, CDMA_0, CDMA_1}. The targeted received power level of macrocell UEs, femtocell UEs and open access UEs are $P$, $Q$, and $P'$, respectively\footnote{We assume a single fixed level of targeted received power at the macrocell or  femtocell level for mathematical tractability. We show that our model is still valid  when the targeted received power is randomly distributed through simulations in Section \ref{section_numerical}.}.  Given the targeted received power $P_T$ ($P_T=P$, $P_T=Q$, or $P_T=P'$) at $\by$  and transmitter at $\bx$, the transmission power is $P_T A|\bx-\by|^{\gamma}$. Then, the resultant interference at $\by'$ is
$\frac{P_T |\bx-\by|^{\gamma}h_{\bx,\by'}}{|\bx-\by'|^{\gamma}}$.

Let $\rho=P'/P$, which is the targeted received power enhancement if a macrocell UE becomes an open access UE. In this paper, we study the performance variation when open access is applied to replace closed access. Therefore, as a parameter corresponding to open access UEs, $\rho$ is regarded as an important designed parameter. Other parameters, such as $P,Q$, and $\gamma$ are considered as predetermined system-level constants.


\subsection{Outage Performance}
In this paper, the performance of macrocell UEs and femtocell UEs (under  open access or closed access) is examined through the outage probability, which is defined as
 the probability  that the signal to interference ratio (SIR) is smaller than a given threshold value $T$. Because we focus on the interference analysis,  the thermal noise is assumed to be negligible in this paper.


\subsection{Scope of This Work} \label{subsection_scope}

The above model assumes a single shared channel for all UEs.  However, the model is  applicable for  the orthogonal multiplexing case (e.g., OFDMA) \cite{SG_MultiTier}. In that case, the spectrum is partitioned into $n$ orthogonal resource blocks, and thus the density of UEs is equivalently reduced by a factor of $n$ when we assume random access of each resource block.

 In this case, $\overline{\nu}=\int_{\mathcal{B}(\bzero, R)}\nu(\bx)d\bx$ is the average number of local femtocell UEs inside a femtocell sharing the same resource block, and $\overline{\lambda}=\pi R^2 \lambda$ is the  average number of open access UEs inside a femtocell sharing the same resource block (in the open access case only).

\section{Open Access vs. Closed Access at the Macrocell Level}\label{section_tier1}

In this section, we analyze the uplink interference and outage performance of macrocell UEs. Consider a reference macrocell UE, termed the \textit{typical  UE}, communicating with its macrocell BS, termed the \textit{typical BS}. 
We aim to investigate the performance of the typical UE.

Due to stationarity of point processes corresponding to macrocell UEs, femtocell BSs, and femtocell
UEs, throughout this section we will re-define the coordinates
so that the typical  BS is located at  $\bzero$ \cite{SG_Totorial1}.
Correspondingly, the  typical UE is located at some $\bx_U$ that is uniformly distributed in $\mathcal{H}(\bzero)$, since macrocell BSs form a deterministic hexagonal grid \cite{SG_Totorial1}.

Let $\Phi'$ be the point process of all other macrocell UEs conditioned on the typical UE, which is called the \textit{reduced Palm point process} \cite{SG_Totorial1} with respect to (w.r.t.) $\Phi$.  Because the reduced Palm point process of a PPP has the same distribution as its original PPP,  $\Phi'$ is still a PPP with intensity $\lambda$ \cite{SG_Totorial1}. Therefore, for presentation convenience, we still use $\Phi$ to denote this reduced Palm point process.



\subsection{Open Access Case} \label{subsection_Open}

\subsubsection{Interference Components}
The overall interference in the uplink  has three parts: from  macrocell UEs not inside any femtocell (denoted by $I_1$), from open access UEs (denoted by $I_2$), and from  femtocell UEs (denoted by $I_3$).

$I_1$ can be computed as the sum of interference from each  macrocell UE:
\begin{align}\label{formula_I1}
I_{1}=\sum_{\bx \in \Phi^0} \frac{P |\bx-\mathcal{BS}(\bx)|^{\gamma}h_{\bx,\bzero}}{|\bx|^{\gamma}},
\end{align}
where $\Phi^0$ denotes the points of $\Phi$ not inside any femtocell.

$I_2$  can be computed as the sum of interference from all open access UEs of all femtocells:
\begin{align}\label{formula_I2}
I_{2}=\sum_{\bx_0\in \Theta}\sum_{\bx\in \Omega(\bx_0)}\frac{P'|\bx-\bx_0|^{\gamma}h_{\bx,\bzero}}{|\bx|^{\gamma}}.
\end{align}

$I_3$ can be computed   as the sum of interference from all femtocell UEs of all femtocells:
\begin{align}\label{formula_I3}
I_{3}=\sum_{\bx_0\in \Theta}\sum_{\bx\in \Psi(\bx_0)}\frac{Q|\bx-\bx_0|^{\gamma}h_{\bx,\bzero}}{|\bx|^{\gamma}}.
\end{align}

The overall interference of open access is $I=I_1+I_2+I_3$.

\subsubsection{Laplace Transform of $I$} \label{subsection_theorem1}
In this subsection, we study the Laplace transform of $I$, denoted by $\mathcal{L}_I$, which leads to the following theorem\footnote{For presentation convenience, we omit the variable $s$ in all Laplace transform expressions.}:
\begin{theorem}\label{theorem1}
\begin{align}\label{formula_theorem1}
\nonumber\mathcal{L}_I=&\mathbf{E}\bigg(\prod_{\bx\in\Phi}u(\bx)\bigg)\cdot\mathbf{E}\Bigg[
\prod_{\bx_0\in\Theta}\frac{\mathbf{E}\Big(\prod_{\bx\in\Omega(\bx_0)}v(\bx,\bx_0)\Big)}{\mathbf{E}\Big(\prod_{\bx\in\Omega(\bx_0)}u(\bx)\Big)}\\
&\qquad\qquad\qquad \mathbf{E}\Big(\prod_{\bx\in\Psi(\bx_0)}w(\bx,\bx_0)\Big)\bigg)
\Bigg],
\end{align}
where $u(\bx)=\exp\left(-\frac{sP |\bx-\mathcal{BS}(\bx)|^{\gamma}h_{\bx,0}}{|\bx|^{\gamma}}\right)$, $v(\bx,\bx_0)=\\\exp\left(-\frac{s\rho P|\bx-\bx_0|^{\gamma}h_{\bx,\bzero}}{|\bx|^{\gamma}}\right)$, and  $w(\bx,\bx_0)=\exp\left(-\frac{sQ|\bx-\bx_0|^{\gamma}h_{\bx,\bzero}}{|\bx|^{\gamma}}\right)$.

Proof: See Appendix for the proof.
\end{theorem}


%
%
%
%
%
%


\subsubsection{Numeric Computation of $\mathcal{L}_I$}\label{subsection_compute1}
In this subsection, we present a numeric approach to compute $\mathcal{L}_I$ derived in (\ref{formula_theorem1}),  which will facilitate later comparison between open access and closed access.
 Let $\mathcal{L}_0=\mathbf{E}\left(\prod_{x\in\Phi}u(\bx)\right)$, which is a generating functional corresponding to $\Phi$ \cite{SG_Totorial1, SG_book1}. It can be re-written in a standard integral form as follows:
\begin{align}
\nonumber&\mathcal{L}_0=\exp\Bigg(-\lambda\int\limits_{\mathbb{R}^2}\bigg(1-\int\limits_{\mathbb{R}+}e^{-\frac{sP|\bx-\mathcal{BS}(\bx)|^{\gamma}h}{|\bx|^{\gamma}}} H(dh)\bigg)d\bx\Bigg)\\
\label{formula_sub1}&=\exp\Bigg(-\lambda\int_{\mathbb{R}^2}\frac{\frac{sP|\bx-\mathcal{BS}(\bx)|^{\gamma}}{|\bx|^{\gamma}}}{\frac{sP|\bx-\mathcal{BS}(\bx)|^{\gamma}}{|\bx|^{\gamma}}+1}\Bigg)d\bx.
\end{align}

%

Given the location of a femtocell BS at $\bx_0$, let $\mathcal{W}(\bx_0)=\mathbf{E}\left(\prod\limits_{\bx\in\Psi(\bx_0)}w(\bx,\bx_0)\right)$, which is a generating functional corresponding to $\Psi(\bx_0)$. It can be expressed in a standard form through the Laplace functional of PPP $\Psi(\bx_0)$,
\begin{align}
\label{formula_W}&\mathcal{W}(\bx_0)=\exp\Bigg(-\int\limits_{\mathcal{B}(\bzero, R)}\frac{\frac{sQ|\bx|^{\gamma}}{|\bx+\bx_0|^{\gamma}}}{\frac{sQ|\bx|^{\gamma}}{|\bx+\bx_0|^{\gamma}}+1} \nu(\bx)d\bx\Bigg).
\end{align}

Similarly, let  $\mathcal{V}(\bx_0)=\mathbf{E}\left(\prod_{\bx\in\Omega(\bx_0)}v(\bx,\bx_0)\right)$, and  $\mathcal{U}(\bx_0)=\mathbf{E}\left(\prod_{\bx\in\Omega(\bx_0)}u(\bx)\right)$, we have
\begin{align}
\label{formula_V}&\mathcal{V}(\bx_0)=\exp\Bigg(-\lambda\int\limits_{\mathcal{B}(\bzero, R)}\frac{\frac{s\rho P|\bx|^{\gamma}}{|\bx+\bx_0|^{\gamma}}}{\frac{s\rho P|\bx|^{\gamma}}{|\bx+\bx_0|^{\gamma}}+1} d\bx\Bigg),\\
\label{formula_U}&\mathcal{U}(\bx_0)=\exp\Bigg(-\lambda\int\limits_{\mathcal{B}(\bx_0, R)}\frac{\frac{sP|\bx-\mathcal{BS}(\bx)|^{\gamma}}{|\bx|^{\gamma}}}{\frac{sP|\bx-\mathcal{BS}(\bx)|^{\gamma}}{|\bx|^{\gamma}}+1} d\bx\Bigg).
\end{align}

Let $\mathcal{J}(\bx_0)=\frac{\mathcal{V}(\bx_0)}{\mathcal{U}(\bx_0)}\mathcal{W}(\bx_0)$, which is numerically computable through (\ref{formula_W})-(\ref{formula_U}). Finally, we note that
\begin{align}
\nonumber&\mathbf{E}\Bigg[\prod_{\bx_0\in\Theta}\frac{\mathbf{E}\Big(\prod_{\bx\in\Omega(\bx_0)}v(\bx,\bx_0)\Big)}{\mathbf{E}\Big(\prod_{\bx\in\Omega(\bx_0)}u(\bx)\Big)}\mathbf{E}\Big(\prod_{\bx\in\Psi(\bx_0)}w(\bx,\bx_0)\Big)\bigg)
\Bigg]\\
\nonumber=&\mathbf{E}\Bigg[\prod_{\bx_0\in\Theta}\bigg(\frac{\mathcal{V}(\bx_0)}{\mathcal{U}(\bx_0)}
\mathcal{W}(\bx_0)\bigg)\Bigg]=\mathbf{E}\left(\prod_{\bx_0\in\Theta}\mathcal{J}(\bx_0)\right)\\
\label{formula_sub2}=&\exp\left(-\mu\int_{\mathbb{R}^2}\left(1-\mathcal{J}(\bx_0)\right)d\bx_0\right),
\end{align}
where (\ref{formula_sub2}) is derived from the generating functional with respect to PPP $\Theta$.
Substituting (\ref{formula_sub1}) and (\ref{formula_sub2}) into (\ref{formula_theorem1}), we can numerically compute  $\mathcal{L}_I$:
\begin{align}
\label{formula_LIfinal}\mathcal{L}_I=\mathcal{L}_0\exp\left(-\mu\int_{\mathbb{R}^2}\left(1-\mathcal{J}(\bx_0)\right)d\bx_0\right).
\end{align}

The overall logic to the above is as follows: First, in terms of the Laplace transform, additive interference is in the \emph{product} form, and interference decrease is in the \emph{division} form. Suppose that there are no femtocells at the beginning, and $\mathcal{L}_0$ corresponds to the interference from macrocell UEs.  Then, we add femtocells to the system. Given a femtocell BS at $\bx_0$, $\mathcal{W}(\bx_0)$ corresponds to the interference from  local femtocell UEs inside the femtocell,
$\mathcal{V}(\bx_0)$ corresponds to interference from  open access UEs inside the femtocell, and $\mathcal{U}(\bx_0)$ corresponds to interference \emph{decrease} of open access UEs as they disconnect from their original macrocell BS. Thus, $\mathcal{J}(\bx_0)=\frac{\mathcal{V}(\bx_0)}{\mathcal{U}(\bx_0)}\mathcal{W}(\bx_0)$ represents the overall interference variation when a femtocell centered at $\bx_0$ is added. Finally, $\exp\left(-\mu\int_{\mathbb{R}^2}(1-\mathcal{J}(\bx_0))d\bx_0\right)$ is the overall interference variation after adding all femtocells. As a consequence, the overall interference can be computed in formula (\ref{formula_LIfinal}).

\subsubsection{Outage Probability}\label{subsection_outage1}

Given the SIR threshold $T$,  the outage probability of the typical UE can be computed as the probability that the signal strength $P h_{\bx_U, \bzero}$ over the interference $I$ is less than $T$:
\begin{eqnarray}
\label{formula_Pout} P^o_{out}=\mathbf{P}(P h_{\bx_U, \bzero} < T I)=1-\mathcal{L}_I|_{s=\frac{T}{P}}.
\end{eqnarray}
The last equality above is due to  $h_{\bx_U, \bzero}$ being exponentially distributed with unit mean. As a result, $P^o_{out}$ can be derived directly from  $\mathcal{L}_I$. 

\subsection{Closed Access Case}\label{subsection_Close}
Different from the open access case, the overall interference has only two parts: from  macrocell UEs (denoted by $\widehat{I}_1$) and from  femtocell UEs (denoted by $\widehat{I}_3$).

$\widehat{I}_1$ can be computed as the sum of interference from each  macrocell UE:
\begin{align}\label{formula_I1c}
\widehat{I}_{1}=\sum_{\bx \in \Phi} \frac{P |\bx-\mathcal{BS}(\bx)|^{\gamma}h_{\bx,\bzero}}{|\bx|^{\gamma}}.
\end{align}
$\widehat{I}_3$ is exactly the same as $I_3$ in (\ref{formula_I3}).

Then, the total interference can be computed as $\widehat{I}=\widehat{I}_1+\widehat{I}_3$. Similar to Section \ref{subsection_compute1}, the Laplace transform of $\widehat{I}$ is
\begin{scriptsize}
\begin{align}
\nonumber&\mathcal{L}_{\widehat{I}}=\mathbf{E}\Bigg[\prod_{\bx\in\Phi}u(\bx)\prod_{\bx_0\in\Theta}\prod_{\bx\in\Psi(\bx_0)}w(\bx,\bx_0)\Bigg]\\
\label{formula_closeI}=&\mathcal{L}_0\mathbf{E}\Bigg[\prod_{\bx_0\in\Theta}\bigg(
\mathcal{W}(\bx_0)\bigg)\Bigg]=\mathcal{L}_0\exp\left(-\mu\int_{\mathbb{R}^2}(1-\mathcal{W}(\bx_0))d\bx_0\right),
\end{align}
\end{scriptsize}
where $\mathcal{L}_0$ is the same as (\ref{formula_sub1}), and $\mathcal{W}(\bx_0)$ is the same as (\ref{formula_W}).

The overall logic to the above is as follows: First,   $\mathcal{L}_0$ corresponds to the interference of all macrocell UEs. Given a femtocell BS at $\bx_0$, $\mathcal{W}(\bx_0)$ corresponds to interference from  local femtocell UEs inside the femtocell. Then, $\exp\left(-\mu\int_{\mathbb{R}^2}(1-\mathcal{W}(\bx_0))d\bx_0\right)$ is the overall interference from all femtocells. As a consequence, the overall interference can be computed as formula (\ref{formula_closeI}).

Finally, the outage probability of the typical UE can be computed as
\begin{eqnarray}\label{formula_Pout12}
 P^c_{out}=\mathbf{P}(P h_{\bx_U, \bzero}< T\widehat{I})=1-\mathcal{L}_{\widehat{I}}|_{s=\frac{T}{P}}.
\end{eqnarray}

\subsection{Parameter Normalization}\label{subsection_normalization1}
From the above performance analysis of both open access and closed access, we see that one can can normalize the radius of macrocells $R_c$ to $1$, so that $R$ is equivalent to the ratio of the radius of femtocells to that of macrocells ($R \ll1$). Also, we can normalize the target received power of macrocell UEs $P$ to $1$, so that $Q$ is equivalent to the ratio of the target received power of femtocell UEs to that of macrocell UEs, and $P'=\rho$. Therefore, in the rest of this section, without loss of generality, we set $R_c=1$ and $P=1$.

\subsection{Open Access vs. Closed Access}\label{subsection_OvsC1}
We compare the outage performance of open access and closed access at the macrocell level.
Due to the integral form of the Laplace transform, the expressions of outage probabilities for both open and closed access cases are in non-closed forms, requiring multiple levels of integration. As a consequence, we are motivated to derive  closed-form bounds to compare open access and closed access.

Let $\mathbf{V}_{\max}=4\pi^2R^4(T\rho)^{\frac{2}{\gamma}}\Big(\frac{1}{8}+\frac{1}{4(\gamma+2)}+\frac{1}{(\gamma+2)(\gamma-2)}\Big)$, $\mathbf{V}_{\min}=2\pi^2R^4(T\rho)^{\frac{2}{\gamma}} \Big(\frac{1}{8}+\frac{1}{4(\gamma+2)}+\frac{1}{(\gamma+2)(\gamma-2)}\Big)$, and $C_u$ be a system-level constant predetermined by $T$ and $\gamma$, shown in (\ref{formula_Cu}) of the proof to Theorem \ref{theorem_bound1}.
The closed-form bounds are presented in the following theorem:
\begin{theorem}\label{theorem_bound1}

$-\mathbf{V}_{\max}+\pi R^2 C_u e^{-\overline{\nu}}> 0$ is a sufficient condition for $P^o_{out}< P^c_{out}$, and $-\pi R^2 C_u e^{\overline{\lambda}}+\mathbf{V}_{\min}e^{-\overline{\lambda}-\overline{\nu}}> 0$ is a sufficient condition for $P^c_{out}< P^o_{out}$.

Proof: See Appendix for the proof.
\end{theorem}

Through Theorem \ref{theorem_bound1}, the closed-form expressions can be used to compare the outage probabilities between open access and closed access without the computational complexity introduced by numeric integrations in (\ref{formula_LIfinal}) and  (\ref{formula_closeI}).

In the following, we focus on the performance variation if open access is applied to replace closed access. The parameter corresponding to open access UEs, $\rho$,  is regarded as a designed parameter. If we fix all the other network parameters, increasing $\rho$ implies better performance for open access UEs, but it will also increase the interference from open access UEs to macrocell BSs. As a consequence, we aim to derive $\rho^*$, such that $P^o_{out}=P^c_{out}$. At the macrocell level, macrocell UEs experience less outage iff $\rho<\rho^*$. Thus, $\rho^*$ is referred to as the \emph{maximum power enhancement tolerated at the macrocell level}. Thus, in the deployment of open access femtocells, the network operator is motivated to limit $\rho$ below $\rho^*$ to guarantee that the performance of macrocell UEs under open access is no worse than that under closed access. One way to derive $\rho^*$ is through numerical computation of (\ref{formula_LIfinal}) and  (\ref{formula_closeI}) and numerical search, which introduces high  computational complexity due to the multiple levels of integrations. A more efficient alternative is to find the bounds of $\rho^*$ through  Theorem \ref{theorem_bound1}. Simple algebra manipulation leads to
 \begin{align}
 \label{formula_rhostar1}\rho^*_{\min}=&\frac{1}{T}\left(\frac{C_ue^{-\overline{\nu}}}{4\pi R^2\left(\frac{1}{8}+\frac{1}{4(\gamma+2)}+\frac{1}{(\gamma+2)(\gamma-2)}\right)}\right)^{\frac{\gamma}{2}},\\
 \label{formula_rhostar2} \rho^*_{\max}=&\frac{1}{T}\left(\frac{C_ue^{\overline{\nu}+2\overline{\lambda}}}{2\pi R^2\left(\frac{1}{8}+\frac{1}{4(\gamma+2)}+\frac{1}{(\gamma+2)(\gamma-2)}\right)}\right)^{\frac{\gamma}{2}},
 \end{align}
where  $\rho^*_{\min}$ and $\rho^*_{\max}$ are the lower bound and upper bound of $\rho^*$, respectively. If the network operator limits $\rho<\rho^*_{\min}$, the performance of macrocell UEs under open access  can be guaranteed no worse than their performance under closed access.

Through (\ref{formula_rhostar1}) and (\ref{formula_rhostar2}), we observe that  $\rho^*_{\min}=\mathcal{O}(\frac{1}{R^{\gamma}})$ and $\rho^*_{\max}=\mathcal{O}(\frac{1}{R^{\gamma}})$,  leading to the following corollary:

\begin{corollary}\label{corollary1}
$\rho^*=\mathcal{O}(\frac{1}{R^{\gamma}})$.
\end{corollary}

Intuitively, as a rough estimation,  open access UEs have their distance to the BS reduced approximately by a factor of $R$, leading to the capability to increase their received power by the corresponding gain in the propagation loss function, as their average interference level is maintained. However,  Corollary \ref{corollary1} cannot be trivially obtained from the above intuition. This is because the outage probability does not only depend on the average interference, but also depends on the distribution of the interference (i.e., the Laplace transform of the interference). 
By comparing  (\ref{formula_LIfinal}) with  (\ref{formula_closeI}), if we switch from closed access to open access,  the distribution of the interference will change drastically. Corollary \ref{corollary1} can be derived only after rigorously comparing and bounding the Laplace transforms of interference under open access and closed access.
%
%

Finally, because $\rho^*_{\min}$ and $\rho^*_{\max}$ have the same scaling behavior, Corollary \ref{corollary1} also demonstrates the tightness of the bounds in (\ref{formula_rhostar1}) and (\ref{formula_rhostar2}).

%

\section{Open Access vs. Closed Access at the Femtocell Level}\label{section_tier2}
In this section, we analyze the uplink interference and outage performance of femtocell UEs. Given a reference femtocell UE, termed as the \textit{typical femtocell UE}, connecting with its femtocell BS, termed as the \textit{typical femtocell BS}, we aim to study the interference at the typical femtocell BS. We also define the femtocell corresponding to the typical femtocell BS as the \textit{typical femtocell}, and the macrocell BS nearest to the typical femtocell BS as the \textit{typical macrocell BS}.

Similar to Section \ref{section_tier1},  we re-define the coordinate of the  typical macrocell BS  as $\bzero$. Correspondingly, the  typical femtocell BS is locating at some $\bx_B$ that is uniformly distributed in $\mathcal{H}(\bzero)$ \cite{SG_Totorial1}.
Given the typical femtocell centered at $\bx_B$, let $\Theta'$ denote the point process of other femtocell BSs conditioned on the typical femtocell BS, i.e., the reduced Palm point process w.r.t.~$\Theta$.  Then,  $\Theta'$ is still a PPP with intensity $\mu$ \cite{SG_Totorial1}. For presentation convenience, we  still use $\Theta$ to denote this reduced Palm point process.  Let $\widetilde{\Psi}(\bx_B)$ denote the other femtocell UEs inside the typical femtocell conditioned on the typical femtocell UE. Similarly, $\widetilde{\Psi}(\bx_B)$ has the same distribution as $\Psi(\bx_B)$. Let $\widetilde{\Omega}(\bx_B)$ denote  open access UEs connecting to the typical femtocell BS.


\subsection{Open Access Case} \label{subsection_Open2}

The overall interference in the uplink of the typical femtocell UE has five parts:  from  macrocell UEs not inside any femtocell ($I_1'(\bx_B)$), from open access UEs outside the typical femtocell ($I_2'(\bx_B)$), from femtocell UEs outside the typical femtocell ($I_3'(\bx_B)$), from local femtocell UEs inside the typical femtocell ($I_4'(\bx_B)$), and from open access  UEs inside the typical femtocell ($I_5'(\bx_B)$). We have
\begin{align}
\label{formula_I1b}
I_{1}'(\bx_B)=&\sum_{\bx \in \Phi^0} \frac{P |\bx-\mathcal{BS}(\bx)|^{\gamma}h_{\bx,\bx_B}}{|\bx-\bx_B|^{\gamma}},\\
\label{formula_I2b}
I_{2}'(\bx_B)=&\sum_{\bx_0\in \Theta}\sum_{\bx\in \Omega(\bx_0)}\frac{\rho P|\bx-\bx_0|^{\gamma}h_{\bx,\bx_B}}{|\bx-\bx_B|^{\gamma}},\\
\label{formula_I3b}
I_{3}'(\bx_B)=&\sum_{\bx_0\in \Theta}\sum_{\bx\in \Psi(\bx_0)}\frac{Q|\bx-\bx_0|^{\gamma}h_{\bx,\bx_B}}{|\bx-\bx_B|^{\gamma}},\\
\label{formula_I4b}
I_{4}'(\bx_B)=&\sum_{\bx\in \widetilde{\Psi}(\bx_B)}Q h_{\bx,\bx_B},\\
\label{formula_I5b}
I_{5}'(\bx_B)=&\sum_{\bx\in \widetilde{\Omega}(\bx_B)}\rho P h_{\bx,\bx_B}.
\end{align}
The overall interference is $I'(\bx_B)=\sum_{i=1}^{5}I_i'(\bx_B)$.

%
%
%

Similar to the derivations in Sections \ref{subsection_theorem1} and \ref{subsection_compute1}, the Laplace transform of $I'(\bx_B)$, denoted by $\mathcal{L}_{I'}(\bx_B)$, is  derived as
\begin{align}
\nonumber&\mathcal{L}_{I'}(\bx_B)=\mathcal{L}'_{0}(\bx_B)\exp\Bigg(-\mu\int_{\mathbb{R}^2}1-\\
\label{formula_final2L1}&\quad\frac{\mathcal{V}'(\bx_0,\bx_B)\mathcal{W}'(\bx_0,\bx_B) }{\mathcal{U}'(\bx_0,\bx_B)} d\bx_0\Bigg)\frac{\mathcal{W}''(\bx_B)\mathcal{V}''(\bx_B)}{\mathcal{U}''(\bx_B)},
\end{align}
where
\begin{scriptsize}
\begin{align}
\label{formula_component1}\mathcal{L}_0'(\bx_B)=&\exp\Bigg(-\lambda\int_{\mathbb{R}^2}\frac{\frac{sP|\bx-\mathcal{BS}(\bx)|^{\gamma}}{|\bx-\bx_B|^{\gamma}}}{\frac{sP|\bx-\mathcal{BS}(\bx)|^{\gamma}}{|\bx-\bx_B|^{\gamma}}+1}d\bx\Bigg),\\
\label{formula_WW}\mathcal{W}'(\bx_0,\bx_B)=&\exp\Bigg(-\int\limits_{\mathcal{B}(\bx_0, R)}\frac{\frac{sQ|\bx-\bx_0|^{\gamma}}{|\bx-\bx_B|^{\gamma}}}
{\frac{sQ|\bx-\bx_0|^{\gamma}}{|\bx-\bx_B|^{\gamma}}+1}\nu(\bx-\bx_0)d\bx\Bigg),\\
\label{formula_VV}\mathcal{V}'(\bx_0,\bx_B)=&\exp\Bigg(-\lambda\int\limits_{\mathcal{B}(\bx_0, R)}\frac{\frac{s\rho P|\bx-\bx_0|^{\gamma}}{|\bx-\bx_B|^{\gamma}}}
{\frac{s\rho P|\bx-\bx_0|^{\gamma}}{|\bx-\bx_B|^{\gamma}}+1}d\bx\Bigg),\\
\label{formula_UU}\mathcal{U}'(\bx_0,\bx_B)=&\exp\Bigg(-\lambda\int
\limits_{\mathcal{B}(\bx_0, R)}\frac{\frac{sP|\bx-\mathcal{BS}(\bx)|^{\gamma}}{|\bx-\bx_B|^{\gamma}}}
{\frac{sP|\bx-\mathcal{BS}(\bx)|^{\gamma}}{|\bx-\bx_B|^{\gamma}}+1}d\bx\Bigg),\\
\label{formula_WPP}\mathcal{W}''(\bx_B)=&e^{-\frac{sQ\overline{\nu}}{sQ+1}},
\mathcal{V}''(\bx_B)=e^{-\frac{ s\rho P\overline{\lambda}}{s\rho P+1}},\\
\label{formula_UPP}\mathcal{U}''(\bx_B)=
&\exp\Bigg(-\lambda\int\limits_{\mathcal{B}(\bx_B, R)} \frac{\frac{sP|\bx-\mathcal{BS}(\bx)|^{\gamma}}{|\bx-\bx_B|^{\gamma}}}{\frac{sP|\bx-\mathcal{BS}(\bx)|^{\gamma}}{|\bx-\bx_B|^{\gamma}}+1} d\bx\Bigg).
\end{align}
\end{scriptsize}

%

Similar to (\ref{formula_Pout}), the outage probability (given $\bx_B$) is
\begin{scriptsize}
\begin{align}
\label{formula2_Pout} \widehat{P}_{out}^o(\bx_B)=\mathbf{P}(Q h_{\bx_U, \bx_B}< T I'(\bx_B))=1-\mathcal{L}_{I'}(\bx_B)|_{s=T'},
\end{align}
\end{scriptsize}where $\bx_U$ is the coordinate of the typical femtocell UE (irrelevant to the result), $T'=\frac{T}{Q}$, and  $T$ is the SIR threshold. Because  $\bx_B$ is uniformly distributed in $\mathcal{H}(\bzero)$, the average outage probability can be computed as $\int_{\mathcal{H}(\bzero)}\widehat{P}_{out}^o(\bx_B)d\bx_B/|\mathcal{H}(\bzero)|$, where $|\mathcal{H}(\bzero)|=\frac{3\sqrt{3}R_c^2}{2}$ is the area of a macrocell.

\subsection{Closed Access Case} \label{subsection_Close2}
The overall interference  has three parts: from  macrocell UEs ($\widehat{I}'_1(\bx_B)$), from  femtocell UEs outside the typical femtocell ($\widehat{I}_3'(\bx_B)$), and from  femtocell UEs inside the typical femtocell ($\widehat{I}_4'(\bx_B)$).
$\widehat{I}'_1(\bx_B)$ can be computed as
\begin{align}\label{formula_I1bb}
\widehat{I}_{1}'(\bx_B)=\sum_{\bx \in \Phi} \frac{P |\bx-\mathcal{BS}(\bx)|^{\gamma}h_{\bx,\bx_B}}{|\bx-\bx_B|^{\gamma}},
\end{align}
and $\widehat{I}_3'(\bx_B)$ and $\widehat{I}_4'(\bx_B)$ are exactly the same as $I'_3(\bx_B)$ in (\ref{formula_I3b}) and $I'_4(\bx_B)$ in (\ref{formula_I4b}), respectively.



Thus, the overall interference is $\widehat{I}'(\bx_B)=\widehat{I}_1'(\bx_B)+\widehat{I}_3'(\bx_B)+\widehat{I}_4'(\bx_B)$.
Then, the Laplace transform of $\widehat{I}'(\bx_B)$ is
\begin{align}
\label{formula_final2L2}&\mathcal{L}_{\widehat{I}'}(\bx_B)=\\
\nonumber&\qquad\mathcal{L}'_{0}(\bx_B)\cdot\exp\left(-\mu\int_{\mathbb{R}^2}1-\mathcal{W}'(\bx_0,\bx_B)d\bx_0\right)\cdot\mathcal{W}''(\bx_B).
\end{align}

The outage probability (given $\bx_B$) is
\begin{align}
\label{formula2_Pout2}\widehat{P}_{out}^c(\bx_B)=1-\mathcal{L}_{\widehat{I}'}(\bx_B)|_{s=T'}.
\end{align}

The average outage probability is $\int_{\mathcal{H}(\bzero)}\widehat{P}_{out}^c(\bx_B)d\bx_B/|\mathcal{H}(\bzero)|$.
Similar to the discussion in Section \ref{subsection_normalization1}, we still can normalize $R_c$ and $P$. Hence, in the rest of this section, without loss of generality, we set $R_c=1$ and $P=1$.

\begin{figure*}
\centering  \vspace*{0pt}
\includegraphics[scale=.5]{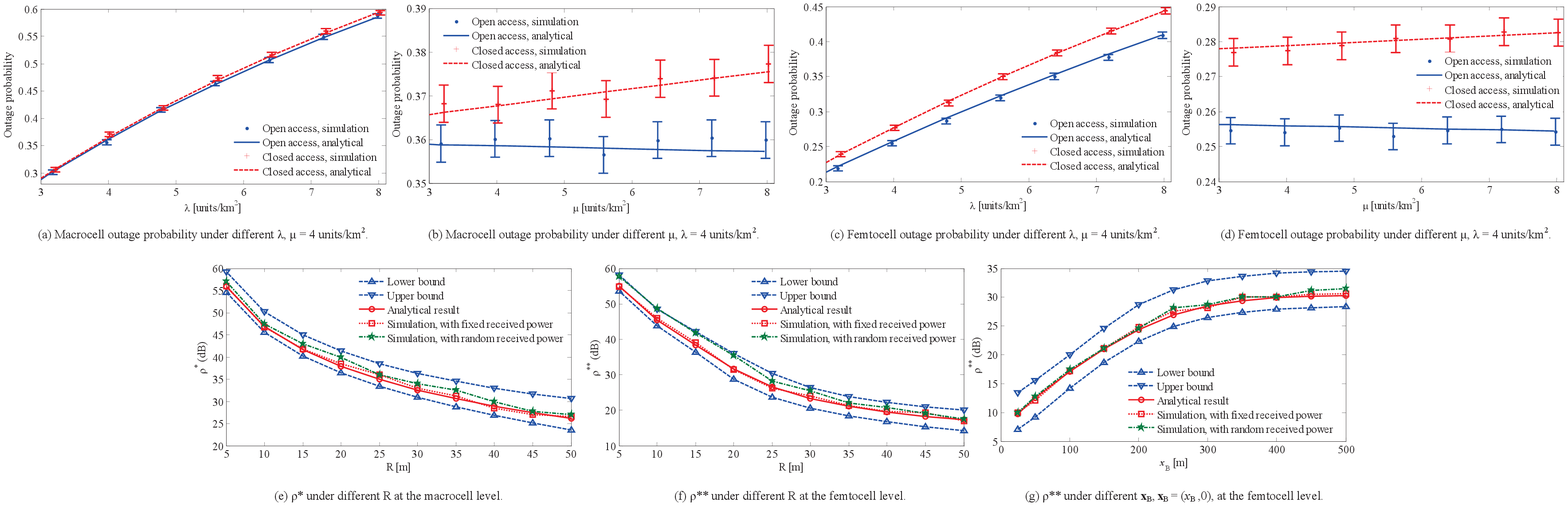}
\caption{Numerical results.}
\label{fig0}
\end{figure*}

\subsection{Open Access vs. Closed Access}\label{subsection_OvsC2}
In this subsection, we compare the outage performance of open access and closed access at the femtocell level.

Let $\mathbf{V}_{\max}'=4\pi^2R^4(T'\rho)^{\frac{2}{\gamma}}\left(\frac{1}{8}+\frac{1}{4(\gamma+2)}+\frac{1}{(\gamma+2)(\gamma-2)}\right)$, $\mathbf{V}_{\min}'\\=2\pi^2R^4(T'\rho)^{\frac{2}{\gamma}}\left(\frac{1}{8}+\frac{1}{4(\gamma+2)}+\frac{1}{(\gamma+2)(\gamma-2)}\right)$;
 $C_u'$ be a system-level parameter predetermined by $T'$ and $\gamma$ similar to $C_u$ in Theorem \ref{theorem_bound1}; $\mathcal{R}_{\min}(\bx_B)$ and $\mathcal{R}_{\max}(\bx_B)$ be as shown in (\ref{formula_Rmin}) and (\ref{formula_Rmax}) in the proof of Theorem \ref{theorem_bound2}, which are in the closed forms if $\gamma$ is a rational number\footnote{It is acceptable to assume $\gamma$ as a rational number in reality, because each real number can be approximated  by a rational number with arbitrary precision.}. Then we have the following theorem:

\begin{theorem}\label{theorem_bound2}
Given $\bx_B$, $K_1\triangleq-\mu \mathbf{V}_{\max}'+\mu\pi R^2 C_u' e^{-\overline{\nu}}-\frac{\pi R^2 T'\rho}{T'\rho+1}+\mathcal{R}_{\min}(\bx_B)>0$ is a sufficient condition for $\widehat{P}^o_{out}(\bx_B)<\widehat{P}^c_{out}(\bx_B)$, and $K_2\triangleq-\mu \pi R^2 C_u'e^{\overline{\lambda}}+\mu\mathbf{V}_{\min}'e^{-\overline{\nu}-\overline{\lambda}}+\frac{\pi R^2 T'\rho}{T'\rho+1}-\mathcal{R}_{\max}(\bx_B)>0$ is a sufficient condition for $\widehat{P}^c_{out}(\bx_B)<\widehat{P}^o_{out}(\bx_B)$.

Proof: See Appendix for the proof.
\end{theorem}

Through Theorem \ref{theorem_bound2}, the closed-form expressions can be used to compare the outage probabilities between open access and closed access without the computational complexity introduced by numeric integrations in (\ref{formula2_Pout}) and (\ref{formula2_Pout2}).

Similar to the discussion in Section \ref{subsection_OvsC1}, let  $\rho^{**}$ denote the value of $\rho$, such that $\widehat{P}^o_{out}(\bx_B)=\widehat{P}^c_{out}(\bx_B)$. At the femtocell level, given that a femtocell BS is located at $\bx_B$ (the relative coordinate w.r.t.~the nearest macrocell), its local femtocell UEs experience less outage iff $\rho<\rho^{**}$. Thus, $\rho^{**}$ is referred to as the \emph{maximum power enhancement tolerated by the femtocell}.

Instead of deriving $\rho^{**}$  through (\ref{formula2_Pout}) and (\ref{formula2_Pout2}), which introduces high computational complexity due to multiple levels of integrations, we can find the lower bound $\rho^{**}_{\min}$ and upper bound $\rho^{**}_{\max}$ of $\rho^{**}$ through Theorem \ref{theorem_bound2}. Accordingly, $\rho^{**}_{\min}$ is the value satisfying $K_1=0$ and $\rho^{**}_{\max}$ is the value satisfying $K_2=0$. Thus,   $\rho^{**}_{\min}$ and  $\rho^{**}_{\max}$  can be found by a numerical search approach w.r.t.~the closed-form expressions.

\section{Numerical Study}\label{section_numerical}

We present simulation and numerical studies on the outage performance in the two-tier network with femtocells.
First, we study the performance of open access and closed access under different user and femtocell densities. Second, we present the numerical results of $\rho^*$ and $\rho^{**}$.
Unless otherwise stated, $R_c=500$ m, $R=50$ m, $\gamma=3$; and fast fading is Rayleigh with unit mean. Each simulation data point is averaged over $50000$ trials. The SIR threshold $T$ is set to $0.1$.


First, we study the performance  under different user and femtocell densities\footnote{As discuss in Section \ref{subsection_scope}, these intensities may already account for the multiplicative factor introduced by orthogonal multiplexing.}. The network parameters are as follows:
  $R_c=500$ m;  $\nu(\bx)=80$ units/km$^2$ if $|\bx|<R$, and $\nu(\bx)=0$ otherwise; $P=-60$ dBm, and $Q=P'=-54$ dBm ($\rho=6$ dB).

Fig. \ref{fig0} (a) and (b) show the uplink outage probability of macrocell UEs under different $\lambda$ and $\mu$ respectively. Fig. \ref{fig0} (c) and (d) show the uplink outage probability of femtocell UEs under different $\lambda$ and $\mu$ respectively. The analytical results are derived from the exact expressions in Sections \ref{subsection_Open}, \ref{subsection_Close}, \ref{subsection_Open2}, and \ref{subsection_Close2}, without applying any bounds.
The error bars show the $95\%$ confidence intervals for simulation results. The plot points are slightly shifted to avoid overlapping error bars for easier inspection.
The figures illustrate the accuracy of our analytical results. In addition, the figures show that the macrocell UE density strongly influences the outage probability of both macrocell and femtocell UEs, while the femtocell density only has a slight influence. 
At the macrocell level, increasing the density of femtocell leads to more proportion of macrocell UEs becoming open access UEs, which gives higher performance gap between open access and closed access. At the femtocell level, the interference is observed at femtocell BSs, and the average number of macrocell UEs in a femtocell becomes a more important factor influencing  the performance gap. 

%

%
%

Next, we present the numerical results of $\rho^{*}$ and $\rho^{**}$. The network parameters are as follows:
   $\lambda=4$ units/km$^2$, $\mu=4$ units/km$^2$;   $\nu(\bx)=20$ units/km$^2$ if $|\bx|<R$, and $\nu(\bx)=0$ otherwise; $P=-60$ dBm, and $Q=-54$ dBm.

Fig.~\ref{fig0} (e) presents the value of $\rho^*$ at the macrocell level. We compute the actual value of $\rho^*$ by numerically searching for the value such that  (\ref{formula_Pout})  is equal to (\ref{formula_Pout12}). Through the closed-form expression in Theorem \ref{theorem_bound1}, we are able to derive the upper and lower bounds of $\rho^*$. Through simulation, we can also search for the value of $\rho^*$ such that the simulated outage probability of open access is equal to that of closed access.
 Furthermore, we also simulate a more realistic scenario, in which each macrocell UE randomly selects a targeted received power level among $0.5P$, $P$, $1.5P$, and $2P$ with equal probability. If a macrocell UE is handed-off to a femtocell, then its targeted received power is multiplied by $\rho$ no matter which power level it has selected.
The figure shows that $\rho^*$ is indeed within the upper bound and the lower bound, and the simulated $\rho^*$ agrees with the analytical  $\rho^*$, validating the correctness of our analysis. Furthermore, this remains the case when the targeted received power is random, indicating the usefulness of our analysis in more practical scenarios.

Figs.~\ref{fig0} (f) and (g) present the value of $\rho^{**}$ at the femtocell level.  Fig.~\ref{fig0} (f) shows $\rho^{**}$ under different $R$ as we fixed $\bx_B=(0,100\textrm{m})$. Fig.~\ref{fig0} (g) shows $\rho^{**}$ under different $\bx_B$ ($\bx_B=(x_B,0)$) as we fixed $R=50$ m. The results show that $\rho^{**}$ is indeed within the upper and lower bounds, and the simulated values of $\rho^{**}$ agree with their analytical values, validating the correctness of our analysis. Furthermore, $\rho^{**}$ decreases in $R$ at a rate slightly faster than that of $\rho^*$, while it increases in $x_B$, until saturating when the femtocell BS is near the macrocell edge.  This quantifies when femtocells are more beneficial as they decrease in size and increase in distance away from the macrocell BS.

%
%
%
%
%
%
%
%

%

\section{Conclusions}\label{section_conclusion}
In this work, we provide a theoretical  framework to analyze the performance difference between  open access and closed access in a two-tier femtocell network. Through establishing a stochastic geometric model, we capture the spatial patterns of different network components. Then, we derive the analytical outage performance of   open access and closed access at the macrocell and femtocell levels. As in most uplink interference analysis, the outage probability expressions are in non-closed forms. Hence, we derive closed-form bounds to compare open access and closed access.  Simulations and numerical studies are conducted, validating the correctness of the analytical model as well as the usefulness of the bounds.


\section*{APPENDIX}

\paragraph{Proof of Theorem \ref{theorem1}}

\begin{figure*}[!htb]
\hrule
\begin{align}
\label{formula_giant1}&\mathcal{L}_I(s)=\mathbf{E}\left(\exp(-sI)\right)
=\mathbf{E}\Bigg[\prod_{\bx\in\Phi^{0}}u(\bx)\prod_{\bx_0\in\Theta}\prod_{\bx\in\Omega(\bx_0)}
v(\bx,\bx_0)\prod_{\bx_0\in\Theta}\prod_{\bx\in\Psi(\bx_0)}w(\bx,\bx_0)\Bigg]\\
\label{formula_giant5}=&\mathbf{E}\Bigg[\mathbf{E}\bigg(\prod_{\bx\in\Phi^0}u(\bx)\bigg|\Theta\bigg)
\mathbf{E}\bigg(\prod_{\bx_0\in\Theta}\prod_{\bx\in\Omega(\bx_0)}v(\bx,\bx_0)\bigg|\Theta\bigg)
\mathbf{E}\bigg(\prod_{\bx_0\in\Theta}\prod_{\bx\in\Psi(\bx_0)}w(\bx,\bx_0)\bigg)\bigg|\Theta\bigg)\Bigg]\\
\label{formula_giant6}=&\mathbf{E}\Bigg[\mathbf{E}\bigg(\prod_{\bx\in\Phi^0}u(\bx)\bigg|\Theta\bigg)
\frac{\mathbf{E}\bigg(\prod\limits_{\bx\in\Phi^1}u(\bx)\bigg|\Theta\bigg)}{\mathbf{E}\bigg(\prod\limits_{\bx\in\Phi^1}u(\bx)\bigg|\Theta\bigg)}
\mathbf{E}\bigg(\prod_{\bx_0\in\Theta}\prod_{\bx\in\Omega(\bx_0)}v(\bx,\bx_0)\bigg|\Theta\bigg)\mathbf{E}\bigg(\prod_{\bx_0\in\Theta}\prod_{\bx\in\Psi(\bx_0)}w(\bx,\bx_0)\bigg|\Theta\bigg)\Bigg]\\
\label{formula_giant7}=&\mathbf{E}\Bigg[\mathbf{E}\bigg(\prod_{\bx\in\Phi}u(\bx)\bigg|\Theta\bigg)
\frac{\mathbf{E}\bigg(\prod_{\bx_0\in\Theta}\prod_{\bx\in\Omega(\bx_0)}v(\bx,\bx_0)\bigg|\Theta\bigg)}
{\mathbf{E}\bigg(\prod_{\bx_0\in\Theta}\prod_{\bx\in\Omega(\bx_0)}u(\bx)\bigg|\Theta\bigg)}
\mathbf{E}\bigg(\prod_{\bx_0\in\Theta}\prod_{\bx\in\Psi(\bx_0)}w(\bx,\bx_0)\bigg|\Theta\bigg)\Bigg]\\
\label{formula_giant9}=&\mathbf{E}\bigg(\prod_{\bx\in\Phi}u(\bx)\bigg)\mathbf{E}\Bigg[
\prod_{\bx_0\in\Theta}\bigg(\frac{\mathbf{E}\Big(\prod_{\bx\in\Omega(\bx_0)}v(\bx,\bx_0)\Big)}{\mathbf{E}\Big(\prod_{\bx\in\Omega(\bx_0)}u(\bx)\Big)}\mathbf{E}\Big(\prod_{\bx\in\Psi(\bx_0)}w(\bx,\bx_0)\Big)\bigg)
\Bigg].
\end{align}
\hrule
\end{figure*}

\begin{proof}
The steps to derive Theorem \ref{theorem1} is shown in (\ref{formula_giant1})-(\ref{formula_giant9}), where  $\Phi^0$ is the point process corresponding to  macrocell UEs not inside any femtocell,  $\Phi^1$ is the point process corresponding to  macrocell UEs inside some femtocell, and $\Phi$ is the aggregation of $\Phi^0$ and $\Phi^1$.

By the law of total expectation, we derive (\ref{formula_giant5}) from (\ref{formula_giant1}).
$\Phi^1$  can be rewritten as the union of all the open access UEs in each femtocell, thus $\mathbf{E}\bigg(\prod_{\bx\in\Phi^1}u(\bx)\bigg|\Theta\bigg)$ is equal  to   $\mathbf{E}\bigg(\prod_{\bx_0\in\Theta}\prod_{\bx\in\Omega(\bx_0)}u(\bx)\bigg|\Theta\bigg)$.
In addition, because $\Phi$ is the aggregation of $\Phi^0$ and $\Phi^1$, $\mathbf{E}\bigg(\prod_{\bx\in\Phi^0}u(\bx)\bigg|\Theta\bigg)$
$\mathbf{E}\bigg(\prod_{\bx\in\Phi^1}u(\bx)\bigg|\Theta\bigg)$ is equal to $\mathbf{E}\bigg(\prod_{\bx\in\Phi}u(\bx)\bigg|\Theta\bigg)$. By considering the two equalities, we derive (\ref{formula_giant7}) from (\ref{formula_giant6}).
Finally, we obtain (\ref{formula_giant9}) from the conditional expectation theorem.
 \end{proof}

\paragraph{Proof of Theorem \ref{theorem_bound1}}

\begin{proof}
In this proof, we use the fact that $P$ and $R_c$ can be normalized and set $P=R_c=1$. Furthermore, we substitute $s=T$ into the integrals in (\ref{formula_V}) and (\ref{formula_U}) to define $\mathcal{V}(\bx_0)=\exp\Bigg(-\lambda\int\limits_{\mathcal{B}(\bzero, R)}\frac{\frac{T\rho |\bx|^{\gamma}}{|\bx+\bx_0|^{\gamma}}}{\frac{T\rho |\bx|^{\gamma}}{|\bx+\bx_0|^{\gamma}}+1} d\bx\Bigg)$ and
$\mathcal{U}(\bx_0)=\exp\Bigg(-\lambda\int\limits_{\mathcal{B}(\bx_0, R)}\frac{\frac{T|\bx-\mathcal{BS}(\bx)|^{\gamma}}{|\bx|^{\gamma}}}{\frac{T|\bx-\mathcal{BS}(\bx)|^{\gamma}}{|\bx|^{\gamma}}+1} d\bx\Bigg).$

\textbf{(a) A sufficient condition for $P^o_{out}< P^c_{out}$}

According to (\ref{formula_LIfinal}), (\ref{formula_Pout}), (\ref{formula_closeI}), and (\ref{formula_Pout12}), $P^o_{out}< P^c_{out}$ iff
\begin{align}
\label{formula_bound11}\frac{\exp\bigg(-\mu\int_{\mathbb{R}^2}\Big(1-\frac{\mathcal{V}(\bx_0)}{\mathcal{U}(\bx_0)}\mathcal{W}(\bx_0)\Big)d\bx_0\bigg)}
{\exp\bigg(-\mu\int_{\mathbb{R}^2}\Big(1-\mathcal{W}(\bx_0)\Big)d\bx_0\bigg)}> 1,
\end{align}
which is equivalent to
\begin{align}
\label{formula_bound12}\int_{\mathbb{R}^2}\left(\frac{\mathcal{V}(\bx_0)}{\mathcal{U}(\bx_0)}-1\right)\mathcal{W}(\bx_0)d\bx_0>0.
\end{align}

Let $V(\bx_0)=\int_{\mathcal{B}(\bx_0,R)}  \frac{ \frac{T\rho|\bx-\bx_0|^{\gamma}}{|\bx|^{\gamma}}}{ \frac{T\rho|\bx-\bx_0|^{\gamma}}{|\bx|^{\gamma}} +1}  d\bx$, and
$U(\bx_0)=\\\int_{\mathcal{B}(\bx_0,R)}\frac{\frac{T|\bx-\mathcal{BS}(\bx)|^{\gamma}}{|\bx|^{\gamma}}}{\frac{T|\bx-\mathcal{BS}(\bx)|^{\gamma}}{|\bx|^{\gamma}}+1} d\bx$. Substitute $V(\bx_0)$ and $U(\bx_0)$  into (\ref{formula_bound12}), we have
\begin{align}
\label{formula_bound13}\int_{\mathbb{R}^2}\left(\frac{\exp(-\lambda V(\bx_0))}{\exp(-\lambda U(\bx_0))}-1\right)\mathcal{W}(\bx_0)d\bx_0> 0.
\end{align}
It is easy to see that the following inequality is a sufficient condition for (\ref{formula_bound13}):
\begin{align}
\label{formula_bound14}\int_{\mathbb{R}^2}\left(-\lambda V(\bx_0)+\lambda U(\bx_0)\right)\mathcal{W}(\bx_0)d\bx_0> 0.
\end{align}

Let $W_{\min}$ and $W_{\max}$ be the lower bound and upper bound of $\mathcal{W}(\bx_0)$, respectively. According to (\ref{formula_W}), $W_{\max}=1$ and  $W_{\min}=e^{-\overline{\nu}}$. Thus, the following is a sufficient condition for (\ref{formula_bound14}):
\begin{align}
\label{formula_bound15}-W_{\max}\int_{\mathbb{R}^2} V(\bx_0)d\bx_0 + W_{\min}\int_{\mathbb{R}^2} U(\bx_0)d\bx_0> 0.
\end{align}

Let $\mathbf{V}=\int_{\mathbb{R}^2}   V(\bx_0)d\bx_0$, we have the following lemma corresponding to the upper and lower bounds of $\mathbf{V}$. Hence, the following  is a sufficient condition for (\ref{formula_bound15}):
\begin{align}
\label{formula_bound16}-W_{\max} \mathbf{V}_{\max} + W_{\min}\int_{\mathbb{R}^2} U(\bx_0)d\bx_0> 0.
\end{align}

\begin{lemma}\label{lemma_V}
 $\mathbf{V}_{\max}=4\pi^2R^4(T\rho)^{\frac{2}{\gamma}}\left(\frac{1}{8}+\frac{1}{4(\gamma+2)}+\frac{1}{(\gamma+2)(\gamma-2)}\right)$, $\mathbf{V}_{\min}=2\pi^2R^4(T\rho)^{\frac{2}{\gamma}}\left(\frac{1}{8}+\frac{1}{4(\gamma+2)}+\frac{1}{(\gamma+2)(\gamma-2)}\right)$, then $\mathbf{V}_{\min}\leq \mathbf{V}\leq \mathbf{V}_{\max}$.

Proof: See the next subsection.
\end{lemma}

In addition, we have
\begin{align}
\nonumber&\int_{\mathbb{R}^2} U(\bx_0)d\bx_0 =\int_{\mathbb{R}^2}\int_{\mathcal{B}(\bx_0,R)}\left(\frac{\frac{T|\bx-\mathcal{BS}(\bx)|^{\gamma}}{|\bx|^{\gamma}}}{\frac{T|\bx-\mathcal{BS}(\bx)|^{\gamma}}{|\bx|^{\gamma}}+1} \right)d\bx d\bx_0\\
\label{formula_BFU} =&\pi R^2 \int_{\mathbb{R}^2}\left(\frac{\frac{T|\bx-\mathcal{BS}(\bx)|^{\gamma}}{|\bx|^{\gamma}}}{\frac{T|\bx-\mathcal{BS}(\bx)|^{\gamma}}{|\bx|^{\gamma}}+1} \right)d\bx=\pi R^2  C_u,
\end{align}
where
\begin{align}\label{formula_Cu}
C_u=\int_{\mathbb{R}^2}\left(\frac{\frac{T|\bx-\mathcal{BS}(\bx)|^{\gamma}}{|\bx|^{\gamma}}}{\frac{T|\bx-\mathcal{BS}(\bx)|^{\gamma}}{|\bx|^{\gamma}}+1} \right)d\bx
 \end{align}
 is only related  to predetermined system-level constants $T$ and $\gamma$.

As a consequence (\ref{formula_bound16}) becomes
\begin{align}
\label{formula_bound17}-W_{\max} \mathbf{V}_{\max} + W_{\min} \pi R^2 C_u> 0.
\end{align}

\textbf{(b) A sufficient condition for $P^o_{out}> P^c_{out}$}

According to (\ref{formula_LIfinal}), (\ref{formula_Pout}), (\ref{formula_closeI}), and (\ref{formula_Pout12}), $P^o_{out}> P^c_{out}$ iff
\begin{align}
\label{formula_bound21}\frac{\exp\bigg(-\mu\int_{\mathbb{R}^2}\Big(1-\mathcal{W}(\bx_0)\Big)d\bx_0\bigg)}
{\exp\bigg(-\mu\int_{\mathbb{R}^2}\Big(1-\frac{\mathcal{V}(\bx_0)}{\mathcal{U}(\bx_0)}\mathcal{W}(\bx_0)\Big)d\bx_0\bigg)}> 1,
\end{align}
Then the following is a sufficient condition for (\ref{formula_bound21}):
\begin{align}
\label{formula_bound24}\int_{\mathbb{R}^2}\left(-\lambda U(\bx_0)+\lambda V(\bx_0)\right)\frac{\mathcal{V}(\bx_0)}{\mathcal{U}(\bx_0)}\mathcal{W}(\bx_0)d\bx_0> 0.
\end{align}
Let $W_{\min}'$ and $W_{\max}'$ be the lower bound and upper bound  of $\frac{\mathcal{V}(\bx_0)}{\mathcal{U}(\bx_0)}\mathcal{W}(\bx_0)$, respectively. According to (\ref{formula_W}), (\ref{formula_V}), and (\ref{formula_U}), $W_{\max}'=\exp\left(\overline{\lambda}\right)$ and  $W_{\min}'=\exp\left(-\overline{\lambda}-\overline{\nu}\right)$.
%
%
Finally, (\ref{formula_bound26}) is a sufficient condition for (\ref{formula_bound24})
\begin{align}
\label{formula_bound26}-W_{\max}' \pi R^2 C_u + W_{\min}'\mathbf{V}_{\min}> 0.
\end{align}
\end{proof}

\paragraph{Proof of Lemma \ref{lemma_V}}
\begin{proof}
\textbf{Upper Bound of $\mathbf{V}$}
\begin{align}
\label{formula_Vmax_check1}\mathbf{V}=&\int_{\mathbb{R}^2} \int_{\mathcal{B}(\bx_0,R)}  \frac{ \frac{T\rho|\bx-\bx_0|^{\gamma}}{|\bx|^{\gamma}}}{ \frac{T\rho|\bx-\bx_0|^{\gamma}}{|\bx|^{\gamma}} +1}  d\bx d\bx_0\\
\nonumber=&\int_{\mathbb{R}^2} \int_{\mathcal{B}(\bx,R)} \frac{ \frac{T\rho|\bx-\bx_0|^{\gamma}}{|\bx|^{\gamma}}}{ \frac{T\rho|\bx-\bx_0|^{\gamma}}{|\bx|^{\gamma}} +1}  d\bx_0 d\bx \\
\label{Vmax_check1}=& \int_{0}^{\infty} 2\pi r_1 \int_{0}^{R} \frac{ \frac{T\rho r_2^{\gamma}}{r_1^{\gamma}}}{ \frac{T\rho r_2^{\gamma}}{r_1^{\gamma}} +1} 2\pi r_2 dr_2 dr_1 \\
\nonumber\leq & \int_{0}^{\infty} 2\pi r_1 \int_{0}^{R} \mathbf{1}(T\rho \frac{r_2^{\gamma}}{r_1^{\gamma}}\geq1) 2\pi r_2  dr_2 dr_1 +\\
\label{Vmax_check2}&\int_{0}^{\infty} 2\pi r_1 \int_{0}^{R} \mathbf{1}(T\rho \frac{r_2^{\gamma}}{r_1^{\gamma}}<1) \frac{T\rho r_2^{\gamma}}{r_1^{\gamma}} 2\pi r_2  dr_2 dr_1 \\
\label{formula_BFVmax}=&4\pi^2R^4(T\rho)^{\frac{2}{\gamma}}\left(\frac{1}{8}+\frac{1}{4(\gamma+2)}+\frac{1}{(\gamma+2)(\gamma-2)}\right).
\end{align}
In (\ref{Vmax_check1}), the integrated item is in the form of $\frac{X}{X+1}$, where $X=\frac{T\rho r_2^{\gamma}}{r_1^{\gamma}}\geq 0$. The bound of the integrated item can be found as follows: if $X\geq 1$, $\frac{1}{2}\leq\frac{X}{X+1}\leq 1$; otherwise, if $X< 1$, $\frac{X}{2}\leq \frac{X}{X+1}\leq X$. Accordingly, we can separate the integration region into $\frac{T\rho r_2^{\gamma}}{r_1^{\gamma}}\geq1$ region and $\frac{T\rho r_2^{\gamma}}{r_1^{\gamma}}<1$ region. As a consequence, the upper bound of (\ref{Vmax_check1}) can be derived as (\ref{Vmax_check2}).

\textbf{Lower Bound of $\mathbf{V}$}

Following a similar approach as above, we have
\begin{align}
\label{Vmax_check1_2}\mathbf{V}
=& \int_{0}^{\infty} 2\pi r_1 \int_{0}^{R} \frac{ \frac{T\rho r_2^{\gamma}}{r_1^{\gamma}}}{ \frac{T\rho r_2^{\gamma}}{r_1^{\gamma}} +1} 2\pi r_2 dr_2 dr_1 \\
\nonumber\geq & \int_{0}^{\infty} 2\pi r_1 \int_{0}^{R} \mathbf{1}(T\rho \frac{r_2^{\gamma}}{r_1^{\gamma}}\geq1) \pi r_2  dr_2 dr_1 +\\
&\int_{0}^{\infty} 2\pi r_1 \int_{0}^{R} \mathbf{1}(T\rho \frac{r_2^{\gamma}}{r_1^{\gamma}}<1) \frac{T\rho r_2^{\gamma}}{r_1^{\gamma}} \pi r_2  dr_2 dr_1 \\
\label{formula_BFVmin}=&2\pi^2R^4(T\rho)^{\frac{2}{\gamma}}\left(\frac{1}{8}+\frac{1}{4(\gamma+2)}+\frac{1}{(\gamma+2)(\gamma-2)}\right).
\end{align}
\end{proof}

\paragraph{Proof of Theorem \ref{theorem_bound2}}
\begin{proof}
In this proof, we use the fact that $P$ and $R_c$ can be normalized and set $P=R_c=1$. Furthermore, we substitute $s=T'$ into the integrals in (\ref{formula_VV})-(\ref{formula_UPP}) to define
$\mathcal{V}'(\bx_0,\bx_B)=\exp\Bigg(-\lambda\int\limits_{\mathcal{B}(\bx_0, R)}\frac{\frac{T'\rho |\bx-\bx_0|^{\gamma}}{|\bx-\bx_B|^{\gamma}}}
{\frac{T'\rho|\bx-\bx_0|^{\gamma}}{|\bx-\bx_B|^{\gamma}}+1}d\bx\Bigg)$,
$\mathcal{U}'(\bx_0,\bx_B)=\exp\Bigg(-\lambda\int
\limits_{\mathcal{B}(\bx_0, R)}\frac{\frac{T'|\bx-\mathcal{BS}(\bx)|^{\gamma}}{|\bx-\bx_B|^{\gamma}}}
{\frac{T'|\bx-\mathcal{BS}(\bx)|^{\gamma}}{|\bx-\bx_B|^{\gamma}}+1}d\bx\Bigg)$,
$\mathcal{V}''(\bx_B)=e^{-\frac{ T'\rho \overline{\lambda}}{T'\rho +1}}$, and
$\mathcal{U}''(\bx_B)=
\exp\Bigg(-\lambda\int\limits_{\mathcal{B}(\bx_B, R)} \frac{\frac{T'|\bx-\mathcal{BS}(\bx)|^{\gamma}}{|\bx-\bx_B|^{\gamma}}}{\frac{T'|\bx-\mathcal{BS}(\bx)|^{\gamma}}{|\bx-\bx_B|^{\gamma}}+1} d\bx\Bigg).$

\textbf{(a) A sufficient condition for $\widehat{P}^o_{out}(\bx_B)< \widehat{P}^c_{out}(\bx_B)$}

According to (\ref{formula_final2L1}), (\ref{formula2_Pout}), (\ref{formula_final2L2}), and (\ref{formula2_Pout2}),  $\widehat{P}^o_{out}(\bx_B)< \widehat{P}^c_{out}(\bx_B)$ iff
\begin{align}
\label{formula_bound31}\frac{\exp\bigg(-\mu\int_{\mathbb{R}^2}\Big(1-\frac{\mathcal{V}'(\bx_0,\bx_B)}{\mathcal{U}'(\bx_0,\bx_B)}\mathcal{W}'(\bx_0,\bx_B)\Big)d\bx_0\bigg)}
{\exp\bigg(-\mu\int_{\mathbb{R}^2}\Big(1-\mathcal{W}'(\bx_0,\bx_B)\Big)d\bx_0\bigg)}\frac{\mathcal{V}''(\bx_B)}{\mathcal{U}''(\bx_B)}> 1.
\end{align}

Let $V'(\bx_0,\bx_B)=\int_{\mathcal{B}(\bx_0,R)}  \frac{ \frac{T'\rho|\bx-\bx_0|^{\gamma}}{|\bx-\bx_B|^{\gamma}}}{ \frac{T'\rho|\bx-\bx_0|^{\gamma}}{|\bx-\bx_B|^{\gamma}} +1}  d\bx$,
$U'(\bx_0,\bx_B)=\\\int_{\mathcal{B}(\bx_0,R)}\left(\frac{\frac{T'|\bx-\mathcal{BS}(\bx)|^{\gamma}}{|\bx-\bx_B|^{\gamma}}}{\frac{T'|\bx-\mathcal{BS}(\bx)|^{\gamma}}{|\bx-\bx_B|^{\gamma}}+1} \right)d\bx$, and $\mathcal{R}(\bx_B)= \int_{\mathcal{B}(\bx_B, R)}\\ \frac{\frac{T'|\bx-\mathcal{BS}(\bx)|^{\gamma}}{|\bx-\bx_B|^{\gamma}}}{\frac{T'|\bx-\mathcal{BS}(\bx)|^{\gamma}}{|\bx-\bx_B|^{\gamma}}+1}d\bx$. Substituting $V'(\bx_0,\bx_B)$, $U'(\bx_0,\bx_B)$ and $\mathcal{R}(\bx_B)$ into (\ref{formula_bound31}), similar to (\ref{formula_bound14}), the following is a sufficient condition for (\ref{formula_bound31}):
\begin{align}
\nonumber&\mu\int_{\mathbb{R}^2}\left(-\lambda V'(\bx_0,\bx_B)+\lambda U(\bx_0,\bx_B)\right)\mathcal{W}'(\bx_0,\bx_B) d\bx_0\\
\label{formula_bound34}&-\frac{\lambda\pi R^2 T'\rho}{T'\rho+1}+\lambda \mathcal{R}(\bx_B)> 0.
\end{align}

Let $W_{\min}''$ and $W_{\max}''$ be the lower bound and upper bound of $\mathcal{W}'(\bx_0,\bx_B)$, respectively. According to (\ref{formula_WW}), $W_{\max}''=1$ and  $W_{\min}''=\exp\left(-\overline{\nu}\right)$. Thus, the following is a sufficient condition for (\ref{formula_bound34}):
\begin{align}
\nonumber&\mu\int_{\mathbb{R}^2}-\left( V'(\bx_0,\bx_B)W''_{\max}+ U(\bx_0,\bx_B)W''_{\min}\right)d\bx_0\\
\label{formula_bound35}&-\frac{\pi R^2 T'\rho}{T'\rho+1}+\mathcal{R}(\bx_B)> 0,
\end{align}
where $\int_{\mathbb{R}^2}V'(\bx_0,\bx_B)d\bx_0=\int_{\mathbb{R}^2} \int_{\mathcal{B}(\bx_0,R)}  \frac{ \frac{T'\rho|\bx-\bx_0|^{\gamma}}{|\bx|^{\gamma}}}{ \frac{T'\rho|\bx-\bx_0|^{\gamma}}{|\bx|^{\gamma}} +1}  d\bx d\bx_0$ is in the same form as (\ref{formula_Vmax_check1}). Thus, by applying Lemma \ref{lemma_V}, we can derive its upper bound and lower bound as $\mathbf{V}_{\max}'$ and $\mathbf{V}_{\min}'$ from (\ref{formula_BFVmax}) and (\ref{formula_BFVmin}), respectively.  Similar to the derivation of (\ref{formula_BFU}),  $\int_{\mathbb{R}^2}U'(\bx_0,\bx_B)= \pi R^2 C_u'$  where
$C_u'=\int_{\mathbb{R}^2}\left(\frac{\frac{T'|\bx-\mathcal{BS}(\bx)|^{\gamma}}{|\bx-\bx_B|^{\gamma}}}{\frac{T'|\bx-\mathcal{BS}(\bx)|^{\gamma}}{|\bx-\bx_B|^{\gamma}}+1} \right)d\bx$ is a constant predetermined by $T'$ and $\gamma$.

In addition, the lower bound $\mathcal{R}_{\min}(\bx_B)$ and the upper bound $\mathcal{R}_{\max}(\bx_B)$  of $\mathcal{R}(\bx_B)$ can be derived as follows:
\begin{align}
\label{formula_Rmin}&\mathcal{R}_{\min}(\bx_B)=\\
\nonumber&\begin{cases}
\pi \int_0^R \frac{\frac{T'(|\bx_B|)^{\gamma}}{r^{\gamma}}r}{\frac{T'(|\bx_B|)^{\gamma}}{r^{\gamma}}+1} dr  &\mbox{if } |\bx_B| \leq R, \\
\pi \int_0^R \frac{\frac{T'(|\bx_B|-R)^{\gamma}}{r^{\gamma}}r}{\frac{T'(|\bx_B|-R)^{\gamma}}{r^{\gamma}}+1} dr
+\pi \int_0^R \frac{\frac{T'(|\bx_B|)^{\gamma}}{r^{\gamma}}r}{\frac{T'(|\bx_B|)^{\gamma}}{r^{\gamma}}+1} dr & \mbox{if }|\bx_B| > R,
\end{cases}
\end{align}
and
\begin{align}
\nonumber\mathcal{R}_{\max}(\bx_B)=&\pi\int_0^R \frac{\frac{T'(|\bx_B|+R)^{\gamma}}{r^{\gamma}}r}{\frac{T'(|\bx_B|+R)^{\gamma}}{r^{\gamma}}+1} dr+\\
\label{formula_Rmax}&\qquad\qquad\pi\int_0^R \frac{\frac{T'(\sqrt{|\bx_B|^2+R^2})^{\gamma}}{r^{\gamma}}r}{\frac{T'(\sqrt{|\bx_B|^2+R^2})^{\gamma}}{r^{\gamma}}+1} dr.
\end{align}

Note that $\int\frac{Br}{r^{\gamma}+B}dr$ is in closed form when $\gamma$ is a rational number. 
Therefore, both $\mathcal{R}_{\min}(\bx_B)$ and  $\mathcal{R}_{\max}(\bx_B)$  are expressed in closed forms.

Finally,  the following is a sufficient condition for (\ref{formula_bound35}):
\begin{align}
\label{formula_bound36}&-\mu \mathbf{V}_{\max}'+\mu\pi R^2 C_u'W_{\min}''-\frac{\pi R^2 T'\rho}{T'\rho+1}+\mathcal{R}_{\min}(\bx_B)>0.
\end{align}

\textbf{(b) A sufficient condition for $\widehat{P}^o_{out}(\bx_B)> \widehat{P}^c_{out}(\bx_B)$}

$\widehat{P}^o_{out}(\bx_B)> \widehat{P}^c_{out}(\bx_B)$ iff
\begin{align}
\nonumber&\mu\int_{\mathbb{R}^2}\left(-\lambda U'(\bx_0,\bx_B)+\lambda V'(\bx_0,\bx_B)\right)\frac{\mathcal{W}'(\bx_0,\bx_B)\mathcal{V}'(\bx_0,\bx_B)}{\mathcal{U}'(\bx_0,\bx_B)}d\bx_0\\
\label{formula_bound44}&+\frac{\lambda\pi R^2 T'\rho}{T'\rho+1}-\lambda\mathcal{R}(\bx_0)> 0.
\end{align}

Let $W_{\min}'''$ and $W_{\max}'''$ be the lower bound and upper bound value of $\frac{\mathcal{W}'(\bx_0,\bx_B)\mathcal{V}'(\bx_0,\bx_B)}{\mathcal{U}'(\bx_0,\bx_B)}$, respectively. According to (\ref{formula_WW})-(\ref{formula_UU}), $W_{\max}'''=\exp\left(\overline{\lambda}\right)$ and  $W_{\min}'''=\exp\left(-\overline{\lambda}-\overline{\nu}\right)$.
%
Then similarly to the derivation of (\ref{formula_bound36}), we see that the following
 is a sufficient condition for (\ref{formula_bound44}):
\begin{align}
\label{formula_bound45}&-\mu \pi R^2 C_u' W_{\max}'''+\mu\mathbf{V}_{\min}'W_{\min}'''+\frac{\pi R^2 T'\rho}{T'\rho+1}-\mathcal{R}_{\max}(\bx_B)>0.
\end{align}
\end{proof}

\scriptsize
\bibliographystyle{abbrv}
\bibliography{baoweiSG}

%
%
\end{document}